\documentclass[prl,twocolumn,showpacs]{revtex4}
\def\comment#1{}

\usepackage{graphicx}
\begin{document}
\title{On the gravitational and electrodynamical stability of  massive nuclear density cores   }
\author{Vladimir Popov$^{1,2}$, Michael Rotondo$^{3}$, Remo Ruffini$^{1,3,4}$ and She-Sheng Xue$^{1,3}$}
\email{ruffini@icra.it}
\affiliation{
$^{1}$%
ICRANet, Piazzale della Repubblica 10, 65122, Pescara, Italy \\
$^{2}$%
Institute of Theoretical and Experimental Physics, 117 218, Moscow, Russia\\
$^{3}$%
Department of Physics and ICRA, University of Rome `Sapienza', Piazzale A.Moro 5,  00185, Rome, Italy\\
$^{4}$%
ICRANet, University of Nice-Sophia Antipolis, 28 avenue de Valrose, 06103 Nice Cedex 2, France}
\received{ \today}

\begin{abstract}
We present a unified treatment of nuclear density cores recovering the classic results for neutral atoms with heavy nuclei having a mass number $A\approx 10^2$--$10^6$ and extrapolating these results to massive nuclear density cores with $A\approx(m_{\rm Planck}/m_n)^3 \sim 10^{57}$. The treatment consists of solving the relativistic Thomas-Fermi equation describing a system of $N_n$ neutrons, $N_p$ protons and $N_e$ electrons in beta decay equilibrium. The $N_p$ protons are distributed at a constant density  within a spherical core of radius $R_c$. 
A new island of stability  is found for 
$A>A_R = 0.039\left(\frac{N_p}{A}\right)^{1/2}\left( \frac{m_{\rm Planck}}{m_n}\right)^3$
 . The Coulomb repulsion, screened by relativistic electrons, is  balanced by  the gravitational self-interaction of the core.
In analogy to heavy nuclei they present, near their surface, an overcritical electric field.  
The  relation between $A$ and $N_p$ is  generalized to an arbitrary value of the mass number,  and the phenomenological relations for $A< 1.5\cdot 10^{2}$ are obtained as a limiting case. 

\end{abstract}

\pacs{97.60.Jd,04.70.-s,21.65.Mn,03.75.Ss,26.60.Dd,31.15.ht}
\maketitle 
The study of neutral atoms with heavy nuclei of mass number $ A\sim10^2-10^6$  is a classic problem of theoretical physics \cite{zeldovich, report}. Special attention has been given to the study of their stability against Coulomb repulsion \cite{greiner82}  and to the existence of electric  fields larger than the critical value 
\begin{eqnarray} 
E_c=m_e^2c^3/e\hbar,  
\label{pn1q}
\end{eqnarray}
near their surfaces \cite{popov}. We have extrapolated these results by numerical integration to the case of massive nuclear density cores: an overall neutral system of neutrons, protons and electrons, in beta decay equilibrium, at nuclear density and having mass numbers $A \sim \left( m_{\rm Planck}/m_n\right)^3$ where  $m_n$ is the neutron mass and $m_{\rm Planck}=(\hbar c/G)^{1/2}$ \cite{rrx200613}.

In this letter we present a unified treatment of heavy nuclei and of  massive nuclear density cores made possible by an explicit solitonic solution of the relativistic Thomas-Fermi equation which governs these phenomena.  We confirm the existence  of overcritical electric fields near the  surface of massive nuclear density cores. The maximum value is  given by $E_{max}=0.95\sqrt{\alpha}m_{\pi}^2c^3/e\hbar$ where as usual $\alpha=e^2/(\hbar c)$ and $m_{\pi}$ is the pion mass. In contrast with the case of heavy nuclei we prove the existence of a new island of stability for mass numbers obeying the condition $A>A_R = 0.039\left(\frac{N_p}{A}\right)^{1/2}\left( \frac{m_{\rm Planck}}{m_n}\right)^3$. 
In this case the equilibrium against Coulomb repulsion is not due to the surface tension of strong interactions. It originates from the over-whelming effects of the self-gravitational interaction of the massive  dense core. Finally we obtain a generalized relation between the mass number $A$ and atomic number $N_p$ which  encompasses  previous  phenomenological expressions. 

It is well known \cite{segrebook}  that the stability of nuclei in neutral atoms  is guaranteed by the surface tension ${\mathcal E}_{s}\approx 17.5 \cdot A^{2/3}$ MeV, created by the nuclear forces balancing the Coulomb repulsion ${\mathcal E}_{\rm em}\approx(3/5)e^2 N_p^2/R_c$, where $R_c=r_0 A^{1/3}$ is the nuclear radius and $r_0=1.2\cdot 10^{-13} {\rm cm}\approx 0.85(\hbar/m_\pi c)$. If one assumes 
\begin{eqnarray} 
N_p\simeq \frac{A}{2},  
\label{pn1}
\end{eqnarray}
and a constant proton density $n_p=N_p n_0/ A \approx 0.25 (m_{\pi}c/\hbar)^3$,  where $n_0=A/(4 \pi R_c^3/3)$ is the ordinary nuclear density,  ${\mathcal E}_{s}>{\mathcal E}_{\rm em}$ gives the stability condition  $A<A^*=10^{-1}m_n/ (m_{\pi}\alpha)$  \cite{segrebook}.
If one assumes (see \cite{ruffini}) a more accurate phenomenological expression relating $N_p$ and $A$
\begin{eqnarray}
N_p \simeq  \left[\frac{2}{A}+\frac{3}{200}\frac{1}{A^{1/3}}\right]^{-1}\ ,
\label{zae2}
\end{eqnarray} 
the stability condition is somewhat cumbersome and
corresponds to $A<A^*\approx 1.5 \cdot 10^2$.


A novel situation occurs when super-heavy nuclei ($A > \tilde A\sim 10^4$) are examined \cite{ruffini, rrx200613}. The distribution of electrons penetrates inside the nucleus: a much smaller effective net charge of the nucleus occurs due to the screening of  relativistic electrons \cite{migdal7614, ruffini, ruffinistella81}. A treatment based on the relativistic Thomas-Fermi model has been developed in order to describe  the penetration of the electrons and their effective screening of the positive nuclear charge. In particular, by assuming Eq.~(\ref{pn1}), in a series of classic papers Greiner {\it et al.} \cite{pieper69, muller72, greiner82} and Popov {\it et al.} \cite{popov, zeldovich, migdal7614} were able to solve the non-linear Thomas-Fermi equation. 
It was demonstrated in \cite{migdal7614} that the effective positive nuclear charge is confined to a small layer of thickness $\sim  \hbar/\sqrt{\alpha} m_{\pi  }c$. Correspondingly electric fields of strength much larger than the critical value given by 
Eq.~(\ref{pn1q})  for vacuum polarization at the surface of the core are created. Under these conditions,  however, the creation of electron-positron pairs due to the vacuum polarization process does not occur  because
of the Pauli blocking by 
the degenerate electrons \cite{Pauliblock}. Although the screening of the positive charge of the core increases with increasing $N_p$, the super-heavy nuclei nevertheless remain unstable against the Coulomb repulsion \cite{migdal7614}. 

Here we generalize the classic work of Greiner and Popov to the case $A\approx(m_{\rm Planck}/m_n)^3 \sim 10^{57}$. We have also 
relaxed the condition expressed in Eqs.~(\ref{pn1}) they adopted, by explicitly computing the beta decay equilibrium between neutrons, protons and electrons.
A supercritical field still exists in a shell of thickness $\sim \hbar/\sqrt{\alpha} m_{\pi  }c$ at the core surface, and a charged lepton-baryonic core is surrounded by an oppositely charged  leptonic component. Such massive nuclear density 
cores are globally neutral. We show that they are stable against the Coulomb repulsion of the proton component due to the stabilizing effects of the gravitational self-interactions.

The analytic solution representing a core of degenerate neutrons, protons and electrons  is obtained by assuming its radius to satisfy
\begin{eqnarray} 
R_c=\Delta \frac{\hbar}{m_\pi c}N_p ^{1/3}\,,  
\label{pn}
\end{eqnarray}
where $\Delta$ is a parameter such that $\Delta \approx 1$ ($\Delta < 1$) corresponds to nuclear (supranuclear) densities  when applied to ordinary nuclei. The overall Coulomb potential satisfies the Poisson equation 
\begin{eqnarray}
\nabla^2 V(r)= -4\pi e\left[n_p(r)-n_e(r)\right],
\label{eposs}
\end{eqnarray}
with the boundary conditions $V(\infty)=0$ 
(due to the global charge neutrality of the system)
and finiteness of $V(0)$. The  density $n_e(r)$ of the electrons of mass $m_e$ and charge $-e$ is determined by the Fermi energy condition 
on their Fermi momentum $P_e^{F}$
\begin{eqnarray}
E_e^F =  [(P_e^Fc)^2+m_e^2c^4]^{1/2}-m_ec^2 -e V(r)=0\,,
\label{efe}
\end{eqnarray} 
which leads to 
\begin{eqnarray}
n_e(r) = \frac{(P_e^{F})^3}{3\pi^2\hbar^3}=\frac {1}{3\pi^2\hbar^3c^3}\left[e^2V^2(r)+ 2m_ec^2eV(r)\right]^{3/2}. \nonumber\\
\label{elnd}
\end{eqnarray}
By introducing the dimensionless quantities $ x= r/[\hbar/m_{\pi}c]$, $ x_c= R_c/[\hbar/m_{\pi}c]$  and $\chi/r=eV(r)/c\hbar$, the relativistic Thomas-Fermi equation takes the form
\begin{eqnarray}
\frac{1}{3x}  \frac {d^2\chi(x)}{d x^2}
= -\frac{\alpha}{\Delta^3}\theta( x_c- x)
+ \frac {4\alpha}{9\pi}\left[\frac {\chi^2(x)}{ x^2}
+2\frac{m_e}{m_\pi}\frac{\chi}{x}\right]^{3/2}, \nonumber\\  
\label{eqless}
\end{eqnarray}
where   $\chi(0)=0 , \chi(\infty)=0$.
The neutron density $n_n(r)$ is determined by the Fermi energy condition on their Fermi momentum $P_n^{F}$ imposed by beta decay equilibrium
\begin{eqnarray}
E_n^F &=& [(P_n^Fc)^2+m^2_nc^4]^{1/2}-m_nc^2 \nonumber\\
&=& [(P_p^Fc)^2+m^2_pc^4]^{1/2}-m_pc^2 + eV(r), 
\label{npeq1e}
\end{eqnarray} 
which in turn is related to the proton and electron densities by Eqs.~(\ref{eposs}), (\ref{elnd}) and (\ref{eqless}).
These equations have been integrated numerically \cite{rrx200613}.

\begin{figure}[t]
\begin{center}
\includegraphics[width=\hsize,clip]{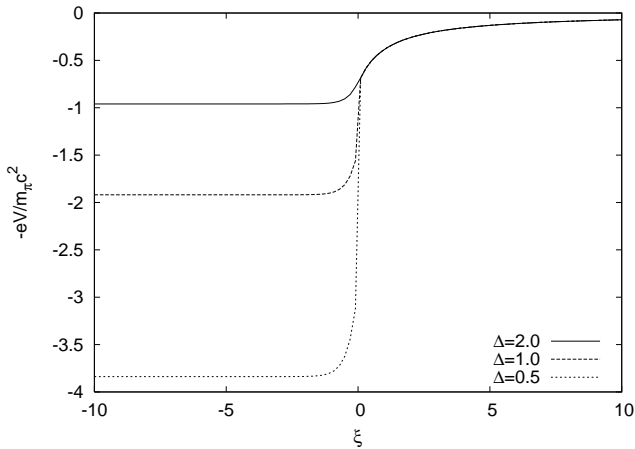}
\end{center}
\caption{The electron Coulomb potential energy $-eV$, in units of pion mass $m_{\pi}$ is plotted as a function of the radial coordinate $\xi=\hat x- \hat x_c$, for selected values of the density parameter $\Delta$.}
\label{efieldf}
\end{figure}

\begin{figure}[t]
\begin{center}
\includegraphics[width=\hsize,clip]{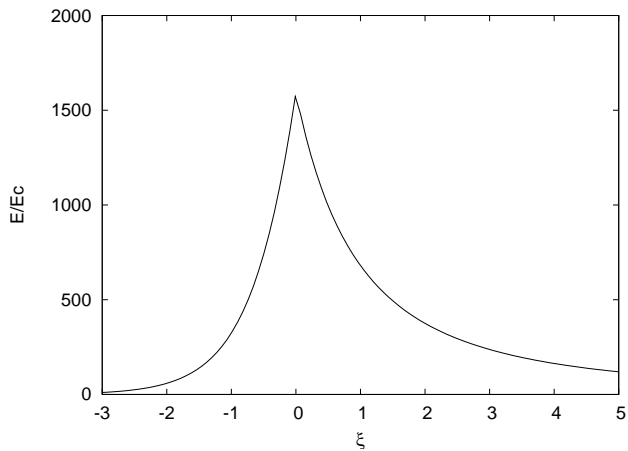}
\end{center}
\caption{The electric field is plotted in units of the critical field $E_c$  as a function of the radial coordinate $\xi$ for $\Delta$=2, showing a sharp peak at the core radius.}
\label{efieldf1}
\end{figure}

In the ultrarelativistic limit,  the relativistic Thomas-Fermi equation admits an analytic solution. Introducing the new function $\phi$ defined by
\begin{eqnarray}\label{anas}
\phi=\Delta\left[\frac{4}{9 \pi}\right]^{1/3}\frac{\chi}{x}\,,
\nonumber
\end{eqnarray}
and the new variables
$\hat x=\left(12/\pi\right)^{1/6}\sqrt{\alpha}\Delta^{-1}x$, $\xi=\hat x- \hat x_c$,  
where $\hat x_c=\left(12/\pi\right)^{1/6}\sqrt{\alpha}\Delta^{-1}x_c$,
then Eq.~(\ref{eqless})  becomes
\begin{eqnarray}
\frac {d^2\hat \phi(\xi)}{d \xi ^2}=-\theta(-\xi)+\hat \phi(\xi)^3\,,
\label{eqless5}
\end{eqnarray}
where  $\hat \phi(\xi)=\phi(\xi+\hat x_c)$.
The   boundary conditions on $\hat \phi$ are: $\hat \phi(\xi)\rightarrow 1$  as $\xi\rightarrow -\hat x_c\ll 0$ 
(at the massive nuclear density core center) and  $\hat \phi(\xi) \rightarrow 0$  as $\xi\rightarrow \infty$. The function $\hat \phi$ and  its first derivative $\hat \phi'$ must be continuous at the surface $\xi=0$ of the massive nuclear density core. 
Eq.~(\ref{eqless5}) admits an exact solution
\begin{eqnarray}
\hat \phi(\xi) 
= \left\{\begin{array}{ll} 1-3\left[1+2^{-1/2}\sinh(a-\sqrt{3}\xi)\right]^{-1}\,, 
&  \xi<0, \\
\displaystyle \frac {\sqrt{2}}{(\xi+b)}\,, & \xi>0\,, 
\end{array}\right.
\label{popovs2}
\end{eqnarray}
where the integration constants $a$ and $b$ have the values $a={\rm arcsinh}(11\sqrt{2})=3.439$, $b=(4/3)\sqrt{2}=1.886$.
We can next evaluate the Coulomb potential energy function
\begin{eqnarray}
eV(\xi)=\left(\frac{9\pi}{4}\right)^{1/3}\frac{1}{\Delta} m_\pi c^2 \hat \phi(\xi)\,,
\label{v01}
\end{eqnarray}
and by differentiation, the electric field
\begin{eqnarray}
E(\xi)=\left(\frac{3^5\pi}{4}\right)^{1/6}\frac{\sqrt{\alpha}}{\Delta^2}\frac{m_\pi^2 c^3}{e\hbar }  \hat \phi'(\xi).
\label{v01e}
\end{eqnarray}
Details  are given in Figs.~\ref{efieldf} and \ref{efieldf1}.

\begin{figure}[th]
\begin{center}
	\includegraphics[width=\hsize,clip]{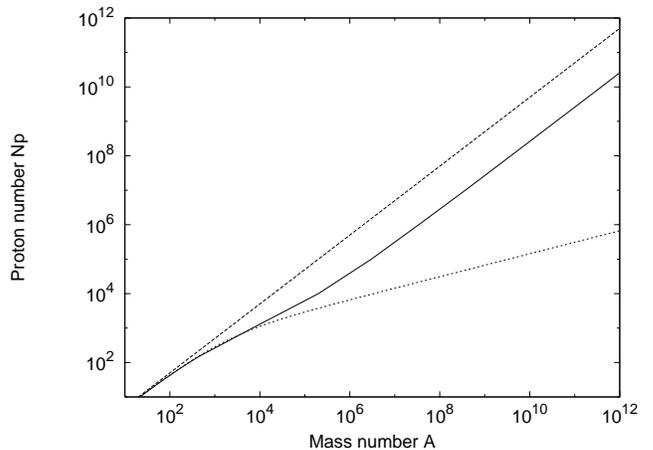}
\end{center}
\caption{ Our $A$-$N_p$ relation at nuclear density (solid line) obtained from first principles compared with the  phenomenological expressions given by Eqs.~(\ref {pn1})  (dashed line) and (\ref{zae2}) (dotted line). The asymptotic value, for $A\rightarrow (m_{\rm Planck}/m_n)^3$, is $N_p\approx A/38$ (see \cite{ruffini20000}).  
}%
\label{ANp}
\end{figure}

Next we  can estimate two crucial quantities: the Coulomb potential at the center of the configuration and the electric field at the surface of the core
\begin{eqnarray}
eV(0)\approx \left(\frac{9\pi}{4}\right)^{1/3} \frac{1}{\Delta} m_\pi c^2\ ,
\label{v0}
\end{eqnarray}
\begin{eqnarray}
E_{\rm max}\approx 0.95\sqrt{\alpha}\frac{1}{\Delta^2}\frac{m_{\pi}^2c^3}{e\hbar}
=0.95\frac{\sqrt{\alpha}}{\Delta^2}\left(\frac{m_{\pi}}{m_e}\right)^2E_c\,. 
\label{v0w}
\end{eqnarray}
Remarkably these two quantities are functions only of the pion mass $m_{\pi}$, the density parameter $\Delta$ and of course the fine constant structure $\alpha$. Their formulas apply over the entire range from superheavy nuclei with $N_p \sim 10^3$ all the way up to massive cores with $N_p \approx (m_{\rm Planck}/m_n)^3 $.

Using the solution (\ref{popovs2}), we have obtained a new generalized relation between $A$ and $N_p$ for any value of $A$ which agrees remarkably well with the phenomenological relations given by Eqs.~(\ref{pn1}) and (\ref{zae2}) in the limit $A< A^*$, as clearly shown in Fig.~\ref{ANp}. 

Having established the validity of the above equations both for superheavy nuclei and for massive nuclear density cores, we now outline some fundamental differences between these two systems. The charge-to-mass ratio 
of the effective charge $Q$ at the core surface to the core mass $M$
is given by
\begin{equation}
\label{neweqn}
  \frac{Q}{\sqrt{G}M} \approx \frac{E_{\rm max}R_c^2}{\sqrt{G}m_n A}
    \approx
  \frac{m_{\rm Planck}}{m_n} 
  \left(\frac{1}{N_p}\right)^{1/3}
  \frac{N_p}{A} \,.
\end{equation}

For superheavy nuclei with $N_p \approx 10^3$ , the charge-to-mass ratio for the nucleus is 
\begin{equation}
\label{neweqn12}
  \frac{Q}{\sqrt{G}M} 
    > \frac{1}{20} \frac{m_{\rm Planck}}{m_n}\sim 10^{18}.
\end{equation} 
There is obviously no role in the stabilization of these nuclei by the gravitational interactions. 

Instead for massive nuclear density cores where $N_p\approx (m_{\rm Planck}/m_n)^3$, the ratio $Q/\sqrt{G}M$ given by Eq.~(\ref{neweqn}) is simply 
\begin{equation}
\label{neweqn1}
  \frac{Q}{\sqrt{G}M} 
    \approx
  \frac{N_p}{A},
\end{equation} 
which is  approximatively $1/38$ (see Fig. \ref{ANp}). It is well-known that the condition that the charge-to-mass-ratio (\ref{neweqn1}) 
be smaller than $1$ is a necessary one for the equilibrium of self-gravitating mass-charge system both in Newtonian and general relativity (see, e.g., \cite{einst}).
Thus massive nuclear density cores are gravitationally stable, globally neutral and bound. It is therefore possible to formulate for them a consistent stable model in terms of gravitational, strong, electromagnetic and weak interactions and quantum statistics \cite{jorge}.

We can see the gravitational stability of massive nuclear density cores from a different point of view.
The {\it maximum} Coulomb energy per proton is given by Eq.(\ref{v0}) where the potential is evaluated at the center of the core.
The gravitational potential energy per proton (of mass $m_p$) in the field of a massive nuclear density core with  $A\approx (m_{\rm Planck}/m_n)^3$, is given by

\begin{eqnarray}
{\mathcal E}_{g} &=&- G\frac{Mm_p}{R_c}=-\frac{1}{\Delta}\frac{m_{\rm Planck}}{m_n}\frac{m_{\pi}c^2}{N_p^{1/3}}\simeq -\frac{m_{\pi}c^2}{\Delta}\left(\frac{A}{N_p}\right)^{1/3}. \nonumber\\
\label{ge1}
\end{eqnarray}
Since $A/N_p \sim 38$ (see Fig. \ref{ANp} ), independently of $\Delta$ value, the gravitational energy is larger in magnitude than and opposite in sign to  the Coulomb potential energy per proton of Eq. (\ref{v0}) so the system is gravitationally stable.

There is yet a third more accurate derivation of the gravitational stability based on the analytic solution of the Thomas-Fermi equation Eq.~(\ref{eqless5}).
The Coulomb energy ${\mathcal E}_{\rm em}$ is \cite{migdal7614} 
\begin{eqnarray}
{\mathcal E}_{\rm em} &=& \int  \frac{E^{2}}{8\pi}d^3r 
\approx  \frac{R_c^2(eV(0))^3 }{(3\pi\alpha)^{1/2}} \int_{-\infty}^{+\infty}\left[\hat\phi'(\xi)\right]^2d\xi \nonumber\\
&=&0.15\frac{3\hbar c(3\pi)^{1/2}}{4\Delta \sqrt{\alpha}}A^{2/3}\frac{m_\pi c}{\hbar}\left(\frac{N_p}{A}\right)^{2/3},
\label{een1}
\end{eqnarray}
which is mainly distributed within a thin shell of width $\delta R_c \approx \hbar \Delta/(\sqrt{\alpha}m_\pi c)$ 
and proton number $\delta N_p= n_p 4\pi R_c^2\delta R_c$ at the surface. 
To ensure the stability of the system,
the attractive gravitational energy of the thin proton shell  
\begin{eqnarray}
{\mathcal E}_{\rm gr}&=&-G\frac{Mm_p \delta N_p}{R_c}\nonumber\\
&\approx &-3\frac{G}{\Delta}
\frac{A^{4/3}}{\sqrt{\alpha}}
\left(\frac{N_p}{A}\right)^{1/3}m_n^2\frac{m_\pi c}{\hbar},
\end{eqnarray}
has to be larger than the repulsive Coulomb energy 
(\ref{een1}). For small $A$ the gravitational energy is always negligible.
However, since the gravitational  energy increases proportionally to $A^{4/3}$ and the Coulomb energy proportionally to $A ^{2/3}$, such a crossing necessarily exists. 
We obtain the crossing at
\begin{eqnarray}\label{sa}
A_R &=& 0.039\left(\frac{N_p}{A}\right)^{1/2}\left( \frac{m_{\rm Planck}}{m_n}\right)^3. \nonumber\\
\end{eqnarray}
This establishes a lower limit for the mass number $A_R$ necessary for the existence of an island of stability for massive nuclear density cores.

Thus  the arguments often quoted, concerning limits on the electric fields of an astrophysical system based on a free test particle approximation 
given by equations like
\begin{eqnarray}
 (E_{\rm max})_{\rm dust}&\approx & \frac{m_e}{e}\frac{m_n c^3}{\hbar}\frac{m_n}{m_{\rm Planck}},
\label{duste}\\
 \left(\frac{Q}{\sqrt{G}M}\right)_{\rm dust} &\approx& \sqrt{G}\frac{m_e}{e}=\frac{1}{\sqrt{\alpha}}\frac{m_e}{m_{\rm Planck}},
 \label{dustq}
\end{eqnarray}
appear to be inapplicable for $A\sim (m_{Planck}/m_n)^3$, when  the collective effects of the quantum statistics are present and  properly taken into account through the relativistic Thomas-Fermi model. Eqs.~(\ref{duste}) and (\ref{dustq}) have to be replaced by Eqs.~(\ref{v0w}) and (\ref{neweqn1}),
\begin{eqnarray}
 E_{\rm max}&=& \frac{0.95\sqrt{\alpha}}{\Delta^2}\frac{ 
m_{\rm Planck}}{m_e}\left(\frac{m_\pi}{m_n}\right)^2(E_{\rm max})_{\rm dust},\label{ede}\\
\frac{Q}{\sqrt{G}M}&=&\frac{N_p}{A}\sqrt{\alpha}\frac{ 
m_{\rm Planck}}{m_e}\left(\frac{Q}{\sqrt{G}M}\right)_{\rm dust}.
 \label{qdq}
\end{eqnarray} 

Having established the role of gravity in stabilizing the Coulomb interaction of the massive nuclear density core, we outline the importance of the strong interactions in determining its surface. We find for the neutron pressure at the surface:
\begin{eqnarray}
P_n
&=&\frac{9}{40}\left(\frac{3}{2\pi}\right)^{1/3}
\left(\frac{m_\pi}{m_n}\right) \frac{m_\pi c^2}{(\hbar/m_\pi c)^3} \left(\frac{A}{N_p}\right)^{5/3}\frac{1}{\Delta^5}, \nonumber\\
\label{prep}
\end{eqnarray}
and for the surface tension, as extrapolated from nuclear scattering experiments,
\begin{equation}
P_s =-\left(\frac{0.13}{4\pi}\right)
\frac{m_\pi c^2}{(\hbar/m_\pi c)^3} \left(\frac{A}{N_p}\right)^{2/3}\frac{1}{\Delta^2}.
\label{surep}
\end{equation}
We then obtain
\begin{eqnarray}
\frac{|P_s|}{P_n} &=&0.39\cdot\Delta^3
\left(\frac{N_p}{A}\right)=0.24\cdot \frac{\rho_{\rm nucl}}{\rho_{\rm surf}},
\label{spr}
\end{eqnarray}
where $\rho_{\rm nucl}=m_n n_0$.
The relative importance of the nuclear pressure and nuclear tension is a very sensitive function of the density at the surface $\rho_{\rm surf}$, 
which in the present case is also constant throughout the configuration. This ratio will in turn determine the extent of a ``crust'' surrounding the core,  
with leptonic pressure and with density due to nuclei.

In conclusion, we have obtained the following results.

a) We have found a new island of  stability in addition to the known one when the  Coulomb repulsion of  heavy nuclei is balanced by the surface tension of strong interactions \cite{segrebook} . The new island occurs for much larger mass numbers: for $A> A_R$ as given by Eq. (\ref{sa}). The Coulomb repulsion is now balanced by the gravitational forces and the role of strong interactions is only relevant in balancing the neutron pressure and determining the size of the core.

b) The systematic use of the relativistic Thomas-Fermi equation and the enforcement of beta equilibrium has allowed to obtain, from first principle, a relation between the mass number $A$ and the proton number $N_p$ both in the mass range of stable nuclei and  in the massive nuclear density cores  for $A\approx(m_{\rm Planck}/m_n)^3 \sim 10^{57}$. For $A<A^*$ this new relation approaches asymptotically the phenomenological relations expressed by  Eqs. (\ref{pn1}) and (\ref{zae2}). For $A>A_R$ it leads to $N_p/A\sim 1/38$. 

c) Although the configurations are  globally neutral, electric fields, whose order of magnitude is larger than the critical field given by Eq. (\ref{pn1q}),  are confirmed to exist close to the surface of the core. The charge to mass ratio at the surface, as well as the values of the electric field  are of the order of $10^{14}$ larger than the ones expected from simpler dust approximations.

We are currently relaxing the constant proton density condition adopted in this article taking also into  account the general relativistic effects by a generalized Tolmann-Oppenheimer-Volkoff equation \cite{OV39} in order to describe the gravitational interactions of such cores \cite{jorge}. The applications of these results to the physics of neutron stars appear to be particularly promising \cite{ruffini20, ruffini200}.
We are also exploring the effects of the  electromagnetic structure of these massive nuclear density cores during the process of gravitational collapse to a Black Hole.

\end{document}